# Optimizing National Security Strategies through LLM-Driven Artificial Intelligence Integration

Dmitry I. Mikhailov, *Senior Member, IEEE*

*Abstract*—As artificial intelligence and machine learning continue to advance, we must understand their strategic importance in national security. This paper focuses on unique AI applications in the military, emphasizes strategic imperatives for success, and aims to rekindle excitement about AI's role in national security. We will examine the United States progress in AI and ML from a military standpoint, discuss the importance of securing these technologies from adversaries, and explore the challenges and risks associated with their integration. Finally, we will highlight the strategic significance of AI to national security and a set of strategic imperatives for military leaders and policymakers.

## I. Introduction

SINCE the early days of cyber space technology development, the United States has made significant strides in enhancing its strategic capabilities. Today, we find ourselves at the precipice of a new technological revolution: Artificial Intelligence (AI). As a strategic imperative for national security, AI presents unparalleled opportunities for strengthening our defense capabilities, similar to how space and cyberspace technology transformed our approach to warfare and reconnaissance.

These technologies have the potential to revolutionize military operations, serving as force multipliers that augment existing capabilities and enable the development of novel operational concepts. Consequently, it is essential for military leaders and policy makers to recognize the strategic importance of AI and integrate them into our planning and decision-making processes.

Although advancements in AI have transformed various sectors of modern society, including business, finance, and production, the national importance of AI as a strategy is still not being adequately reflected in the published strategies of different branches of government. Despite the memorandum issued by Deputy Secretary of Defense Kathleen Hicks in May 2021, which directed the Department of Defense's holistic, integrated, and disciplined approach to Responsible Artificial Intelligence (RAI), many military leaders have yet to integrate AI strategies into their decision-making processes.

Many of these documents, such as the aforementioned memorandum, focus more on the responsible AI aspects, which while important, might not focus some interesting ways we can leverage AI as a fighting force. This paper focuses on unique AI applications in the military, emphasizes strategic imperatives for success, and aims to rekindle excitement about AI's role in national security.

We have intentionally kept the discussion non-technical in nature and maintained a higher-level perspective, despite addressing a technically-oriented readership. This approach allows us to focus on the broader strategic implications and challenges of AI integration in military operations, rather than delving into specific technical details. By adopting this higher-level perspective, we aim to foster a more inclusive conversation that encourages interdisciplinary dialogue and promotes a holistic understanding of the complex issues surrounding AI and its military applications.

## II. Understanding the Technology

Artificial Intelligence is revolutionizing the way military and government organizations operate. These advanced technologies enable machines to learn and reason autonomously, with applications ranging from situational awareness to decision-making support. In particular, the advent of Large Language Models (LLMs) has significantly impacted the field of natural language processing, providing valuable insights from unstructured text data and facilitating human-like communication.

While there are models available, for battlefield use it is more advantageous to develop and train a custom LLM, based on something such as GPT4. This model would be specifically designed and trained on military and government-related data, ensuring a higher level of domain expertise, accuracy, and relevancy in the generated text. Once trained, it can be used for various applications such as intelligence analysis, automated report generation, and natural language interfaces for command and control systems.

Deploying AI systems in battlespace contexts requires careful consideration of three main components: the model, data, and computing environment. The model serves as the digital brain, trained to perform specific tasks such as object

---

recognition, threat prediction, or sentiment analysis. Organizations can either utilize pre-trained models like GPT4 or develop their own custom models tailored to their unique requirements.

Without data, an AI cannot operate. High-quality, labeled data is essential for training and testing models, ensuring that they can generalize well to new, unseen situations. In a military context, this may involve collecting data from various sources, including satellite imagery, communication intercepts, and field reports.

The computing environment is where the AI model runs, and it is crucial to ensure operational security and data integrity. For illustration sake, we suppose an integration flow that involves TensorFlow or PyTorch as the underlying AI framework and then hosting the trained model on secure cloud-based services tailored to government use, such as AWS GovCloud, Microsoft Azure Government, or Google Cloud for Government. We can then deploy the model on SIPRNet (Secret Internet Protocol Router Network) to enable classified and secure use.

Deployment means integrating the AI model into an organization's existing systems, such as a drone's software or a command center's communication platform. For instance, the previously theorized LLM could be integrated into a drone's control system to automatically analyze incoming data and generate mission-critical insights in real-time.

Using DevSecOps and standard pipelines for the deployment of the AI is key, as this will ensure smooth integration, maintenance, and updating of AI models within complex systems. For battlespace usage, it is critical to conduct a comprehensive assessment of the capabilities and limitations of AI technologies to ensure their seamless incorporation into strategic operations and tactical planning.

By carefully considering the model, data, and computing environment, and employing a well-planned integration flow, the military can deploy AI systems that enhance their operational effectiveness, making them better equipped to address the challenges of the modern battlefield.

### III. APPLICATIONS WITHIN MILITARY OPERATIONS

AI and ML technologies have the potential to transform various aspects of military operations, and already we have seen the increase in AI generated malware based on a given description of the target system.

While many of the tools employed by the USAF and other branches may leverage components of ML, they have yet to fully exploit the potential uses of AI. Below are just a few examples of how AI can revolutionize our fighting forces:

1) **Acquisition and Sustainment**. The acquisition and sustainment of military equipment and supplies is a complex and resource-intensive process, and AI can play a critical role in optimizing and streamlining this process. For example, the US Air Force Material Command can use AI algorithms to predict demand and identify the most efficient routes for transportation, minimizing both risks and costs associated with logistics operations. AI can assess the needs of frontline units and schedule resupply missions accordingly, taking into account factors such as weather, terrain, and enemy activity. Additionally, AI can be used to automate inventory management, ensuring that the right equipment and supplies are available when needed. This can help reduce waste and improve the overall efficiency of the supply chain, enabling military forces to maintain readiness and effectiveness in the face of evolving threats.

2) **Cyber Operations**. AI-powered cyber defense systems are critical to defending against the growing threat of cyberattacks. These systems use AI algorithms to analyze network traffic patterns in real-time, enabling the detection and mitigation of potential threats before they cause significant damage. Additionally, AI can be used offensively to identify vulnerabilities in enemy networks and launch targeted cyberattacks, such as denial-of-service attacks, or disrupt their communications and infrastructure. For example, the U.S. Cyber Command has used AI algorithms to help thwart election interference and defend against cyber threats to critical infrastructure, such as the power grid and financial systems.

3) **ISR Capabilities**. Intelligence, surveillance, and reconnaissance (ISR) capabilities are critical to military operations. AI can process vast amounts of data from various sources, including SIGINT, IMINT, and HUMINT, to produce actionable intelligence that can inform military decision-making. Project Maven, a U.S. Department of Defense initiative, is a prime example of this application. By leveraging AI, Project Maven can analyze countless hours of drone footage, identify patterns, and pinpoint areas of interest, resulting in improved intelligence gathering and enhanced situational awareness. AI can also be used to enhance ISR capabilities in other ways, such as by analyzing social media data to identify potential threats or analyzing satellite imagery to monitor troop movements and infrastructure.

4) **Electromagnetic Spectrum Operations (EMSO).** The electromagnetic spectrum is a critical domain of warfare, and AI can play a critical role in managing and exploiting it. AI can analyze and interpret electromagnetic signals to identify and locate enemy emitters, such as communication or radar systems. This information can be used to develop countermeasures, such as jamming or deceiving enemy communications, and manage the electromagnetic spectrum more effectively to maintain friendly forces' superiority in the electronic battle-space. For instance, the U.S. Air Force is exploring the use of AI to optimize its use of the electromagnetic spectrum, including allocating frequencies for communication and sensing, and to enable more effective use of electronic warfare.

5) **Counter-Unmanned Aerial Systems (C-UAS)**. The proliferation of commercial and military drones presents new challenges for military forces. AI can be employed to detect, track, and counteract unauthorized or hostile unmanned aerial systems (UAS). For instance, the U.S. Army's Rapid Equipping Force (REF) has developed the Mobile-Low, Slow, Small Unmanned Aerial Vehicle Integrated Defeat System (M-LIDS), which uses AI algorithms to detect and track low-flying drones and provide countermeasure options to disable or destroy them. This system can enhance the protection of military installations, personnel, and equipment from the potential threats posed by unauthorized or hostile UAS.
6) **Joint Training Exercises.** Joint training exercises are critical to building and maintaining joint force readiness, interoperability, and effectiveness. AI-driven simulations can replicate complex conflict scenarios, providing valuable insights into potential outcomes and enabling the refinement of strategies and tactics. These simulations take into account various factors, such as terrain, weather, and enemy capabilities, to create realistic and dynamic scenarios that challenge military planners and decision-makers. The U.S. Army's Synthetic Training Environment program is exploring the use of AI and virtual reality to create realistic training scenarios that improve soldiers' skills and readiness. Additionally, joint training exercises can also involve live-fire training, which provides opportunities for forces to work together in realistic combat scenarios and improve their interoperability and effectiveness.
7) **Dynamic Targeting**. The U.S. military's Kill Chain model emphasizes the need for rapid and accurate targeting of enemy assets in modern warfare. AI can play a crucial role in expediting the kill chain process by automating target identification, tracking, and prioritization. For instance, the Defense Advanced Research Projects Agency's (DARPA) Collaborative Operations in Denied Environment (CODE) program aims to develop AI-driven autonomous systems that can collaborate and adapt to dynamic combat situations, allowing for faster and more precise targeting of enemy assets.
8) **Submarine Warfare**. AI has the potential to revolutionize undersea warfare by enhancing the capabilities of submarines and other underwater systems. For example, the U.S. Navy's Large Displacement Unmanned Undersea Vehicle (LDUUV) program aims to develop AI-driven autonomous underwater vehicles capable of performing various missions, such as intelligence, surveillance, and reconnaissance (ISR), mine countermeasures, and offensive operations. AI can also be used to improve the effectiveness of submarine communication systems, such as the Extremely Low Frequency (ELF) and Very Low Frequency (VLF) systems, which are essential for maintaining command and control while submerged.

AI technology is continuously advancing at an astonishing pace, leading to innovative integrations and capabilities that are emerging daily. Some applications may seem futuristic, but recent developments showcase the potential impact of AI in the near term.

In April of 2023, government focused data analytics company, Palantir, conducted a demonstration highlighting the potential of generative AI in tactical operations. Using advanced AI models like FLAN-T5 XL, GPT-NeoX-20B, and Dolly-v2-12b LLMs, an operator receives an alert regarding enemy activity and consults an AI chatbot for further intelligence and potential courses of action. The AI chatbot then provides pertinent information and proposes various tactical options, such as deploying an F-16, utilizing long-range artillery, or launching Javelin missiles. Palantir's system streamlines and automates numerous aspects of warfare, with operators primarily seeking guidance from the chatbot and approving its suggestions.

However, as we increasingly integrate LLMs into military operations, it is crucial to swiftly address and mitigate the inherent challenges and risks associated with their deployment. For instance, LLMs are prone to "hallucinating" or fabricating information, which could have dire consequences in the field.

Moreover, we must account for the unique vulnerabilities AI integration presents, such as the potential for adversarial exploitation, to ensure a reliable and secure AI-driven future for our armed forces. By proactively tackling these challenges, we can harness the full potential of AI in advancing military capabilities and national security

IV. SAFEGUARDING OUR COMPETITIVE EDGE

The integration of AI into military operations, while presenting significant advantages, is not without inherent risks. It is crucial to safeguard AI technologies against adversaries seeking to undermine our capabilities through cyberattacks targeting our AI deployments. A multi-faceted approach is essential for protecting sensitive information while still harnessing the benefits of AI advancements. The following principles form the foundation of securing and deploying AI in a manner that mitigates mission risks associated with implementation. This list is not hierarchical or exhaustive but serves as a starting point for strategists aiming to incorporate AI as a core capability within our military forces:
1) **Federated Learning.** Federated learning enables collaborative AI model training across multiple devices or organizations while preserving data privacy. By sharing only model updates and not raw data, federated learning reduces the risk of data leakage and ensures that sensitive information remains secure.
2) **Robust Adversarial Training.** Adversarial machine learning is a technique used by adversaries to create malicious input data designed to deceive AI algorithms, leading to incorrect predictions or classifications. By incorporating adversarial examples into the training process, robust adversarial training helps AI models become more resilient against attacks.
3) **Differential Privacy.** Differential privacy is a technique that adds carefully calibrated noise to data or query results to protect individual data points' privacy. By employing differential privacy, we can prevent adversaries from extracting sensitive information through model inversion attacks, which aim to reveal training data from the model's output.
4) **Secure Enclaves.** Secure enclaves are protected areas within a processor that prevent unauthorized access to data and code execution. By deploying AI models within secure enclaves, we can protect them from attacks such as memory probing, which attempts to extract sensitive information from a model's internal memory.
5) **Model Watermarking.** Model watermarking embeds unique, imperceptible signatures within AI models, enabling their origin and ownership to be traced. This technique can be used to detect model theft or unauthorized use, helping to protect intellectual property and ensure the integrity of AI systems.

6) **Continuous Monitoring and Validation.** Regularly monitoring the performance and behavior of AI can help identify potential security threats, such as data poisoning or Trojan attacks. By continuously validating the input-output relationships of the models, we can detect and mitigate attempts to compromise their integrity and effectiveness.
7) **Red Teaming and Penetration Testing.** Conducting red teaming exercises and penetration testing of AI and ML architectures can help identify potential vulnerabilities and weaknesses. By proactively addressing these issues, we can ensure that our AI technologies remain secure and effective in the face of evolving threats.

By understanding and addressing the challenges of AI integration, we can better harness their strategic and tactical potential, ensuring our nation remains at the forefront of technological advancements. A comprehensive and actionable roadmap for the successful adoption and implementation of AI technologies in the military domain should focus on specific applications, risks, and solutions, allowing us to maintain our competitive edge while safeguarding our AI infrastructure.

## V. THE WAY FORWARD

The integration of AI into military operations presents an immense opportunity for asymmetric advantage. To fully harness this potential, military leaders and policymakers must adopt a comprehensive approach that addresses several key imperatives. Taking from delve into these strategic imperatives and discuss how they can be woven together to ensure AI is effectively incorporated into military operations.

First and foremost, maintaining technological superiority and staying at the forefront of advancements in national security contexts relies on accelerating AI and ML development and research. This involves not only prioritizing investment in AI research but also fostering close partnerships with academia, industry, and international partners. By doing so, the military can ensure that cutting-edge AI technologies are continually incorporated into its applications.

In conjunction with technological advancements, it is crucial to establish ethical, legal, and societal guidelines for AI's responsible use. By defining and enforcing boundaries that align with national values and principles, the military can guarantee that AI technologies are used ethically and legally in all contexts. This approach will also help foster international cooperation, which is essential for the successful deployment of AI technologies in a global security landscape.

The development of a comprehensive AI strategy is a natural extension of the ethical considerations surrounding AI. By formulating a national cyber strategy that encompasses clear objectives, milestones, and metrics, the military can create a roadmap for the development and deployment of AI technologies. This ensures that the military remains at the cutting edge of innovation while also promoting AI resilience and interoperability across different branches of the military.

Addressing resilience and interoperability, the implementation of adversarial training methods can improve the robustness of AI and ML systems. This reduces vulnerability to malicious input data and ensures seamless integration, paving the way for creating a seamless multi-domain AI framework. Developing an AI framework that integrates across all domains of military operations, including land, sea, air, space, and cyberspace, will enable greater interoperability and collaboration.

With an AI framework in place, it becomes essential to cultivate a skilled AI workforce to support the integration of AI and ML technologies. Investing in education, training, and recruitment programs that focus on building a strong foundation in AI and ML skills among military personnel is a crucial step. A diverse and talented workforce is vital for ensuring that AI technologies can be effectively deployed in national security operations.

Building on the foundation of a skilled workforce, incorporating AI capabilities into military decision-making processes can greatly enhance situational awareness and decision-making in complex and dynamic environments. Utilizing AI for real-time data analysis and processing enables military leaders to make better-informed decisions, maintaining strategic advantage in future conflicts. This focus on decision-making is further supported by integrating AI-driven simulations and war games into military training and exercises, increasing mission readiness and operational effectiveness.

As the final piece of the puzzle, adapting military processes, concepts, and doctrines to accommodate AI perspectives and risks is essential for facilitating the successful adoption and implementation of AI technologies. By updating these processes, the military can ensure that it remains agile and adaptable in an ever-evolving technological landscape, fully harnessing the strategic potential of AI in its operations.

## VI. CONCLUSION

As we reflect upon our accomplishments in space and cyber capabilities, we must recognize the immense potential AI holds for our fighting forces and national security. The integration of AI into our defense infrastructure will help us to tackle complex challenges with greater efficiency, accuracy, and speed. It is essential to develop a comprehensive AI strategy that addresses the urgency and intricacy of modern warfare, fosters collaboration between the military, academia, and industry, and ensures immediate developmental follow-up to maintain our position as a foremost leader in military capabilities.

The lessons learned from our experiences with cyber and space development can inform our approach to AI integration. Our focus should be on creating robust, accurate, secure, and ethically responsible AI systems capable of adapting to the dynamic nature of warfare. We must prioritize research, development, and testing of AI technologies that enhance situational awareness, decision-making, and autonomous operations while safeguarding against potential vulnerabilities.

In conclusion, AI stands as a strategic imperative for our national security, just as space and cyber technology did in the past. By leveraging the lessons from these advancements, we can effectively harness the power of AI to maintain our strategic edge, protect our nation's interests, and ensure a safer and more secure future. The integration of AI into our defense capabilities will not only revolutionize the way we conduct operations but also serve as a testament to our unwavering commitment to technological innovation in the pursuit of peace and security.

## ACKNOWLEDGMENT

The author would like to express his sincere gratitude to his mentors and friends, whose support and guidance have been instrumental in shaping his career and in the development of this paper. Special thanks go to Rob Gilliom, Tim Willison, and Col. John Wright, USAF (ret.), whose insights and perspectives have been invaluable. The author would also like to thank Anthony, Sam, and many others whose guidance and encouragement have been much appreciated.

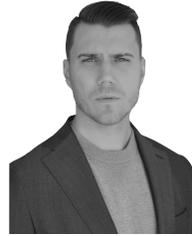

**Dmitry I. Mikhailov** (Senior Member, IEEE) is an accomplished cybersecurity researcher and consultant, with over 12 years of experience in the field. He holds several certifications, including CISM, CASP+, and GPEN, and is currently studying Computer Science at the University of Hertfordshire. He has worked as a cybersecurity engineer in the defense sector, serving in key roles at companies such as Lockheed Martin and SAIC. He currently provides expert consultation to organizations across the globe, helping them solve complex electronic security problems and implement effective strategies.

In addition to his professional experience, Dmitry has contributed to a variety of cybersecurity research and development projects. He designed and developed a Photovoltaic Random Number Generator and contributed to the first updated Linux hardening guide published by SANs in 2011.

Mr. Mikhailov is an active member of several professional societies, including the IEEE Computer Society, the ACM Technology Policy Council, and the IEEE Standards Association. He has received numerous awards and recognitions for his work, including the Presidential Volunteer Service Award and the Congressional Recognition Award. Dmitry currently serves as the Region 5 coordinator for the IEEE Computer Society and is passionate about advancing the field of cybersecurity through his research, publications, and leadership.